\documentclass[twocolumn]{aastex631}

\usepackage{amsmath}
\usepackage{mathtools}
\usepackage{multirow}
\usepackage{tabularx}
\usepackage{natbib}
\usepackage{float}
\usepackage{afterpage}
\usepackage{graphicx}
\usepackage{booktabs}
\usepackage{rotating}
\usepackage{threeparttable}

\newcommand{\Msun}{M_{\odot}}
\newcommand{\Mst}{M_*}

\begin{document}

\title{Illuminating M82: Simulating X-ray Emission from Galactic Winds in a Starburst Galaxy}

\author[0009-0001-2631-6295]{Allison Lin}
\affiliation{Department of Astronomy, Columbia University, New York, NY 10027, USA}
\affiliation{Department of Physics and Astronomy, Vanderbilt University,
2301 Vanderbilt Place, Nashville, TN 37235, USA}
\email{allison.lin@vanderbilt.edu}

\author[0000-0002-0041-4356]{Lachlan Lancaster}
\affiliation{Center for Computational Astrophysics, Flatiron Institute, 162 5th Avenue, New York, NY 10010, USA}
\affiliation{Department of Astronomy, Columbia University, New York, NY 10027, USA}

\author[0000-0002-2499-9205]{Viraj Pandya}
\altaffiliation{NASA Hubble Fellow}
\affiliation{Department of Astronomy, Columbia University, New York, NY 10027, USA}

\author[0000-0003-3806-8548]{Drummond B. Fielding}
\affiliation{Department of Physics, New York University, 726 Broadway, New York, NY, 10003, USA}

\author[0000-0003-2630-9228]{Greg L. Bryan}
\affiliation{Department of Astronomy, Columbia University, New York, NY 10027, USA}

\author[0000-0003-3175-2347]{John A. ZuHone}
\affiliation{Center for Astrophysics $\vert$ Harvard \& Smithsonian, 60 Garden St. Cambridge, MA 02138, USA}

\author[0000-0002-2644-0077]{Sebastian Lopez}
\affiliation{Department of Astronomy, The Ohio State University, 140 W. 18th Ave., Columbus, OH 43210, USA}
\affil{Center for Cosmology and AstroParticle Physics, The Ohio State University, 191 W. Woodruff Ave., Columbus, OH 43210, USA}

\author[0000-0002-1790-3148]{Laura A. Lopez}
\affil{Department of Astronomy, The Ohio State University, 140 W. 18th Ave., Columbus, OH 43210, USA}
\affil{Center for Cosmology and AstroParticle Physics, The Ohio State University, 191 W. Woodruff Ave., Columbus, OH 43210, USA}

\author[0000-0002-5840-0424]{Christopher Carr}
\affiliation{Department of Astrophysical Sciences, Princeton University, New Jersey, NJ 08544, USA}

\begin{abstract}

We generate mock X-ray observations from a suite of idealized high-resolution ($\sim\,4$ pc), tall-box ($\sim2\times2\times8$ kpc$^3$) simulations of star formation driven galactic winds in an M82-like system, varying the spatial resolution as well as the strength and distribution of supernova (SN) energy injection. We compare our mock X-ray observations with deep Chandra observations of the hot plasma around M82. While the simulated total X-ray luminosity, $L_X$, increases with resolution and when SNe feedback is spatially distributed, even in the best case scenario, our simulated $L_X$ is a factor of $\sim\,50-100$ lower than observed and the surface brightness profiles of X-ray emission, $S_X(z)$, fall off too quickly with distance from the galaxy. Past results were able to reproduce these observables and we discuss potential simulation differences that could explain this discrepancy. We make the first comparison of the X-ray spectrum of our simulations to observations and find that our simulated spectrum is too soft, with a deficit of hard X-ray photons at $\gtrsim 1\, {\rm keV}$. We discuss how physical processes missing from our simulations and prior work (e.g., thermal conduction and cosmic rays) could help resolve this discrepancy. 

\end{abstract}


\section{Introduction} \label{sec:intro}
Galactic outflows are commonly observed in both nearby and distant star-forming galaxies \citep[e.g.,][]{Martin1999,Rubin2014,Cicone2016,forsterschreiber19,robertsborsani20,weldon24}. These outflows, when driven by star formation, are primarily due to stellar winds and supernovae (SNe), which inject mass, momentum, energy and metals into the surrounding interstellar medium \citep[ISM;][]{Chevalier&Clegg1985,HeckmanTimothy1990,Veilleux2005,Hopkins2012,fielding17b,thompson24}. Detailed observational \citep[e.g.,][]{heckman15,mcquinn19,DiTeodoro2020,Lopez2025} and theoretical \citep[e.g.,][]{creasey13,kim18,FieldingQuataertMartizzi2018,hu19,pandya21,FieldingBryan2022,smith24,steinwandel24,Richie2024,TanBretFielding2024} studies of the multi-phase structure of galactic outflows have revealed that, as the hot gas flows outward, it brings along other components such as dust, cold gas, and warm gas \citep[e.g.,][]{Rupke2005, Engelbracht2006, Westmoquette2009, Westmoquetteb2009,Leroy2015,Martini2018}. Models of galaxy formation in a cosmological context have long suggested that these galactic winds are crucial for regulating the numbers, sizes, shapes, and colors of galaxies \citep[e.g.,][]{Somerville&Dave2015, Naab&Ostriker2017, CrainRobeRVAN2023}. In particular, the specific energy of SN-driven winds may be the primary parameter controlling the heating, precipitation and accretion rate of gas around galaxies up to the Milky Way scale \citep[e.g.,][]{fielding17,LiBryan20,pandya20,pandya23,carr23,voit24a,voit24b,pandya26}.

One of the best examples of a galactic outflow is the nearby starburst galaxy M82 \citep{Rieke1980}, which offers a unique laboratory to investigate these powerful winds given its proximity \citep[3.6 Mpc;][]{Freedman1994, Greke2011}, nearly edge-on orientation \citep{McKeith1995}, and strong multiphase wind \citep{Leroy2015}. Numerous studies have explored the outflow of M82 across the electromagnetic spectrum. \citet{Walter2002}, \citet{sALAK2013}, \citet{Beirao2015} traced molecular gas and atomic H\textsc{i} at temperatures below 100 K, while \citet{McKeith1995} and \citet{Westmoquette2009} examined ionized gas via H$\alpha$ emission at around $\sim$10$^{4}$ K. In addition to these multi-wavelength studies, several works have focused specifically on the hot X-ray-emitting gas in M82 \citep{Watson1984, Bregman1995, Strickland1997, Strickland2007, Strickland2009, Ranalli2008, Konami2011}. 

Recently, \citet{Lopez2020} analyzed the temperature and metallicity structure of M82’s outflow using deep imaging and spectra from the Chandra X-ray Observatory. They modeled the X-ray emission as a multi-temperature, optically thin thermal plasma with contributions from a non-thermal component and charge exchange\footnote{Charge exchange (CX) is the process in which an ion strips an electron from a neutral atom. In the context of galactic winds, this process represents the interaction of the hot X-ray gas with neutral, colder phases. Incorporating CX has been found to be necessary to accurately model the soft X-ray emission in star-forming galaxies \citep{Liu2012}}. Their findings indicate that the temperatures and number densities of the warm-hot and hot plasma peak at the ``starbursting ridge'' \citep{lester90} and decrease along the minor axis. This is inconsistent with simple spherical adiabatic expansion and instead suggests significant mass loading and mixing with cooler material. Additionally, non-thermal emission was detected in all regions, contributing $\approx13$\% of the total broad-band ($0.5-7$~keV) X-ray flux, while charge exchange contributes 8–25$\%$ of the total flux. Their analysis of the X-ray surface brightness profile revealed an exponential decline along the minor axis, with the highest emission concentrated near the starburst region. Here we utilize high-resolution idealized simulations of an M82-like starburst galaxy to predict mock X-ray surface brightness profiles and spectra for comparison to the measurements by \citet{Lopez2020}.

This paper is organized as follows: In Section \ref{sec:simulations}, we summarize the M82-like hydrodynamical simulations. In Section \ref{sec:methods} we describe our  methods for generating mock X-ray images and spectra. In Section \ref{sec:results}, we present our results, including a comparison of the simulation predictions and observations. In Section \ref{discussion}, we interpret our results in the context of previous work and potential missing physical processes in our simulations. Finally, in Section \ref{conclusion}, we summarize our main findings.

\section{Simulations}
\label{sec:simulations}

To model the multiphase nature of galactic winds, we use the publicly available \texttt{Athena++} code \citep{Stone2020} to run three-dimensional magneto-hydrodynamical (MHD) simulations with a configuration nearly identical to that of \citet{TanBretFielding2024} (which is itself built upon the framework established in \citealt{FieldingQuataertMartizzi2018}). In particular, we simulate a box of dimension $(L_x, L_y, L_z) = (2048,\ 2048,\ 8192)\,{\rm pc}$ with uniform resolution, $
\Delta x$, that is changed across the simulations suite. We use periodic boundary conditions in $x$ and $y$ and outflow boundary conditions in $z$. The simulations are evolved for approximately 100 Myr, with the precise duration varying slightly between the simulations. 

We initialize an exponential gas disc at the mid-plane in the $z$-direction in accordance with the method described in \citet{TanBretFielding2024}, taking the mid-plane density to be $n=300\ {\rm cm}^{-3}$ and scale height $H=86\ {\rm pc}$. This gas distribution is set up to be in hydrostatic equilibrium with a fixed background gravitational potential meant to account for stars and dark matter.

Before initiating SN feedback, we stir the disc with a stochastic Fourier-space turbulent driving field. The driving is applied to modes with integer wave numbers $10 \le |n_i| \le 16$, where $k_i=2\pi n_i/L_i$. For our fiducial box this corresponds to component wavelengths $128 \lesssim \lambda/{\rm pc} \lesssim 205$. The forcing amplitudes are drawn from Gaussian random variates and scaled so that the shell-integrated forcing spectrum obeys $E_{\rm drive}(k)\propto k^{-\alpha}$, with $\alpha=0$ in the simulations presented here; i.e. the forcing power is approximately flat across the driven band rather than concentrated at a single scale. The driving field is localized to the disc by multiplying the real-space perturbations by $\exp[-(z/z_{\rm turb})^2]$, with $z_{\rm turb}=120\,{\rm pc}$, and is evolved as an Ornstein--Uhlenbeck process with correlation time $t_{\rm corr}=8$ Myr. The mass-weighted mean perturbation is subtracted so that no net momentum is added, and the perturbations are rescaled at each update to inject turbulent energy at a fixed rate. The energy injection rate was chosen so that the root-mean-square turbulent velocity is 10 km~s$^{-1}$.

The simulations include radiative cooling, assumed to be optically thin, and a heating term meant to represent photo-electric heating at low temperatures. Due to the combination of the turbulence-driven density structure and the cooling/heating source terms, our ISM evolves into a multiphase state across a wide temperature range (from $\sim$10 K to $10^8$~K)\footnote{We place a temperature floor in the simulations at $T=10\,{\rm K}$, allowing for colder, denser gas than other similar set-ups (which we will return to in the discussion).}.

Once the turbulence has been allowed to evolve for 60 Myr (7.5 vertical eddy turn over times), we initialize SN-driven feedback in two modes: ``concentrated'' and ``distributed.'' In the concentrated case (which makes up the majority of our simulations), all energy and mass is injected in a sphere of radius $r_{\rm inj} = 20\,{\rm pc}$ at the center of the domain. In the distributed case, energy and mass injection is split between 10 locations which are pseudo-randomly distributed within a $500\,{\rm pc}$ radius of the center of the domain in $x$ and $y$ while being constrained to lie at $z=0$. These positions are pseudo-random in that they are initially randomly distributed and then iteratively moved apart from one another so that they are relatively uniformly spaced in $x$ and $y$. Injection of mass and energy at each of these locations occurs within the same $r_{\rm inj}$ radius.

The energy and mass injected into the simulation domain are determined by assuming an instantaneous starburst of total mass $\Mst$. This star mass is turned into a SN rate using a SN-rate per unit stellar mass, $\dot{\mathcal{N}}_{\rm SN}(t)$ following the FIRE-3 prescription \citep{FIRE3} as described in \citet[][see their Figure 3]{TanBretFielding2024}. For each SN we inject $E_{\rm SN} = 10^{51}\,{\rm erg}$ of thermal energy and $M_{\rm SN} = 8.4\,\Msun$ of mass on to the grid within $r_{\rm inj}$. For the distributed case, these SN injections are spread out evenly between the 10 injection sites. The simulation suite explored in this paper varies the numerical resolution of the simulations, $\Delta x$, the total stellar mass driving feedback, $\Mst$, and whether the feedback is concentrated or distributed. The suite is detailed in Table~\ref{tab:sim_part1}.

Since energy injection in these simulations is modeled after a single burst of star formation, feedback only lasts for $\approx 40\,{\rm Myr}$. Similar simulations usually parameterize their feedback in terms of a constant star formation rate, $\dot{M}_*$. We can relate our stellar masses to a star formation rate by taking the average SN rate during the active feedback stage of our simulations, $\langle \dot{N}_{\rm SN}\rangle$, and comparing it to the asymptotic ($t\to \infty$) SN rate for a continuous starburst as calculated using the FIRE-3 functional form: 
\begin{equation}
    \dot{N}_{\rm SN, lim} = \dot{M}_*
    \lim_{t \to \infty} \mathcal{N}_{\rm SN}(t)  \, ,
\end{equation}
where we assume $\dot{M}_*$ is a constant. Ignoring the very minor contribution of Type-Ia SNe, we evaluate $\dot{N}_{\rm SN, lim}$ at the end of the core-collapse SN stage ($t=44\,{\rm Myr}$) to be $ \dot{N}_{\rm SN, lim} = 1.15\times 10^{-2}\, \dot{M}_*/\Msun$. The average SN rate in our $M_* = 10^8\,\Msun$ ($10^7\,\Msun$) simulations is $\langle \dot{N}_{\rm SN}\rangle = 2.86\times 10^{-2}\, {\rm yr}^{-1}$ ($2.86\times 10^{-3}\, {\rm yr}^{-1}$) meaning that these simulations are roughly equivalent to star formation rates of $\dot{M}_* = 2.47 \, \Msun/{\rm yr}$ ($0.25 \, \Msun/{\rm yr}$).

The implied star formation rate for our $M_* = 10^8\,\Msun$ simulations is low by roughly a factor of $\sim 5$ in comparison to values from recent observational studies in the literature which estimate $\dot{M}_* \approx 12\, \Msun/{\rm yr}$ \citep{Vulic2018,Bolatto24}. This fact is crucial to the interpretation of our comparison to the data presented in Section~\ref{sec:results}. As we will discuss further there, we observe a roughly linear relationship between the total X-ray luminosity, $L_{\rm X}$, and the input SN energy (parameterized by $\Mst$), so that we may still make fair comparisons to the data by extrapolating our results to higher equivalent star formation rates.

\section{Methods}
\label{sec:methods}

We predict the X-ray emission using two different methods: (1) using pre-tabulated X-ray emissivities to estimate the intrinsic brightness, and (2) using a Monte Carlo approach to generate synthetic X-ray spectra and images tailored specifically for the Chandra Advanced CCD Imaging Spectrometer (ACIS) instrument. For the latter we use \texttt{pyXSIM} \citep{ZuHone2016} and \texttt{SOXS} \citep{ZuHone2023}. We have verified that the two are generally consistent in terms of the underlying assigned X-ray emissivity (see Appendix \ref{sec:appendix} for more details). 

Our first method estimates the local X-ray emissivity using the local temperature, density, and chemical abundances. Our simulation does not track chemical evolution, so for simplicity we assume solar metallicity ($Z_{\odot}$) for all fluid elements. This assumption is motivated by observations of M82's disk, which have measured approximately solar metallicity in its stellar and gaseous components \citep{Origlia2004}. We adopt the default solar abundance table used by \texttt{pyXSIM}, which assumes \citet{AndersGrevesse1989} abundances. Then we assign an X-ray emissivity for the 0.5$-$7.0 keV energy range to every fluid element using pre-computed emissivity tables from the CHIANTI atomic database \citep{Yong2016} that assumes the same abundance level as \texttt{pyXSIM}. The X-ray emissivity is given by
\begin{equation}
    \label{eq:xray_emissivity}
    \varepsilon_{X} = n_e n_i \Lambda_X(T, Z_{\odot})
\end{equation}
where $n_e$ is the electron number density,  $n_i$ is the ion number density, $T$ is the fluid temperature, and $\Lambda$ is the X-ray emissivity function interpolated from CHIANTI look-up tables. The total X-ray luminosity is then calculated by calculating $\varepsilon_X$ for each cell in the simulation and integrating over the simulation volume. This calculation is computationally inexpensive but makes many simplifying assumptions about the propagation of X-rays through the galaxy and inter-galactic medium (it assumes they are not absorbed at all). For this reason we only compare these measurements to absorption-corrected values inferred from the observations.

We also use this method to predict the X-ray surface brightness profile by integrating the X-ray emissivity in each of 11 evenly spaced bins along the outflow $z$-direction to match \cite{Lopez2020}. The luminosity in each region is then divided by the projected area of the bin to obtain the surface brightness. We define the radial coordinate $r$ as the bin center position relative to the bin of maximum emission, such that $r=0$ corresponds to the peak of the X-ray emission (the center of the galactic plane) and adjacent bins are spaced by the bin width in $z$.

Our second approach involves generating synthetic X-ray images and spectra, accounting for the Chandra-ACIS instrumental response and absorption/scattering by foreground dust in the Milky Way. Briefly, we first use \texttt{pyXSIM} \citep{ZuHone2016} to generate mock photon ``events'' by sampling from a theoretical X-ray spectrum. This step employs a Monte Carlo method to randomly assign energies, positions, and directions to individual photons based on the local emissivity of the simulated gas, ensuring that denser, hotter regions contribute more emission. We assume purely thermal emission from a plasma in collisional ionization equilibrium, utilizing the APEC \citep{Smith2001} plasma code from AtomDB \citep{Foster2012}. We assume an exposure time of 550 ks, similar to the Chandra observations compiled by \citet{Lopez2020}. Finally, we use \texttt{SOXS} \citep{ZuHone2023} to apply the Chandra-ACIS instrument response function. The generated photons are convolved with the Ancillary Response Function (ARF) and Redistribution Matrix Function (RMF) corresponding to the Chandra ACIS-S Cycle 10 detector\footnote{For simplicity, we do not attempt to match the same sequence of cycles for the specific Chandra ObsIDs compiled by \citep{Lopez2020}. We use Cycle 10 as it approximately corresponds to the bulk of the observations used in \citep{Lopez2020}}. Appendix \ref{sec:appendix} shows that the two methods agree relatively well.

\begin{table}
\centering
\normalsize
\setlength{\tabcolsep}{5.5pt}
\renewcommand{\arraystretch}{1.05}

\begin{tabular}{lccc}
\toprule
\textbf{Simulations} & \textbf{Resolution} & $M_*$ & \textbf{Feedback} \\
\midrule
\texttt{Mc1e7\_LR}            & 8 pc & \(10^{7}M_\odot\) & Concentrated \\
\texttt{Mc1e7\_HR}            & 4 pc & \(10^{7}M_\odot\) & Concentrated \\
\texttt{Mc1e8\_LR}            & 8 pc & \(10^{8}M_\odot\) & Concentrated \\
\texttt{Mc1e8\_HR}            & 4 pc & \(10^{8}M_\odot\) & Concentrated \\
\textbf{\texttt{dis\_Mc1e8\_HR}} & 4 pc & \(10^{8}M_\odot\) & \textbf{Distributed} \\
\bottomrule
\end{tabular}%

\caption{Simulations used in this work. All simulations have a physical box size of 2.048 kpc$\times$2.048 kpc$\times$8.192 kpc. See text for details on simulation parameters.}
\label{tab:sim_part1}
\end{table}

\section{Results}
\label{sec:results}

We first present properties of our fiducial simulation in Section \ref{sec:viz} before investigating the phase-space distribution of its X-ray emission in Section \ref{sec:phase}. We then examine the total X-ray luminosity for all simulations (subsection \ref{sec:timeseries}), present mock Chandra X-ray images and spectra (subsection \ref{sec:mocks}), and finally surface brightness profiles (subsection \ref{sec:profiles}).

\subsection{Temperature, density, and X-ray emissivity maps}
\label{sec:viz}

\begin{figure*} 
    \centering
    \includegraphics[width=1\textwidth]{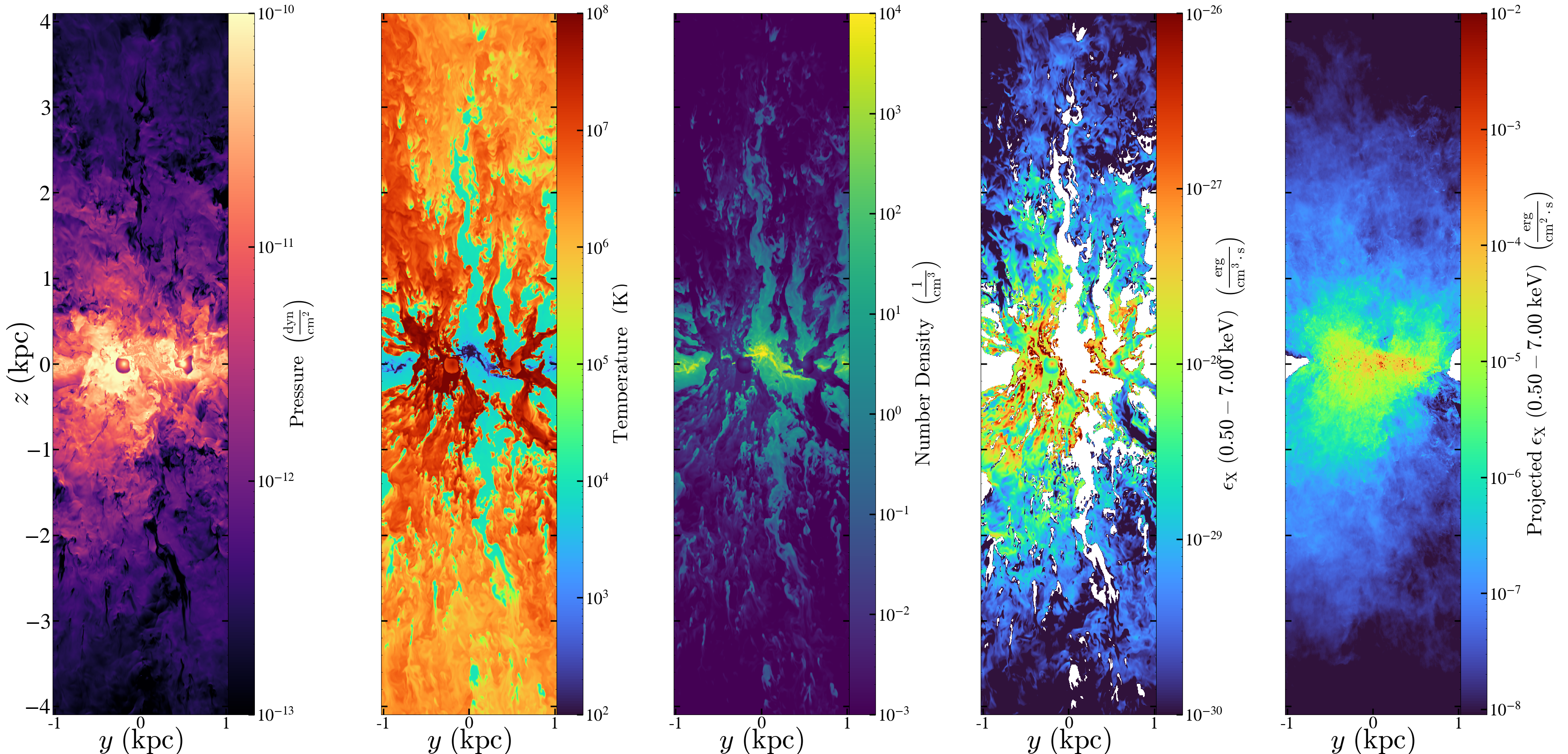}
    \caption{Slices through the center of the simulation in the y--z plane from the highest total emission snapshot of the simulation \texttt{dis\_Mc1e8\_HR}. Left to right: thermal pressure, temperature, number density, intrinsic X-ray emissivity from 0.5-7 keV, and its line-of-sight projection along $x$. The intrinsic X-ray emission arises at the interfaces between the hot wind and the cooler entrained gas, but line-of-sight projection smears these interfaces out, making the emission appear far more volume-filling.}
    \label{fig:slices}
\end{figure*}

To illustrate the spatial distribution of key physical properties, we show slice maps of the pressure, temperature, and number density on the y--z plane, along with slice and projection maps of X-ray emissivity, where the projected X-ray emissivity shows the line-of-sight integrated emissivity along the x-direction. Figure ~\ref{fig:slices} shows these maps for the simulation \texttt{dis\_Mc1e8\_HR} at $t = 90.85 \,{\rm Myr}$. The temperature slice shows the thermal structure of the outflow, highlighting regions of hot gas heated by clustered stellar feedback near the disk, as well as cooler, denser material interspersed throughout the disk and also carried away by the wind. Shock-heated regions and phase interfaces are visible. The slice in X-ray emissivity illustrates how X-ray emission is dominated by interfaces between the hot wind and the surrounding colder gas entrained by the wind \citep{Strickland2000,ToalaArthur18}. The projected X-ray emissivity in the final panel somewhat erases the signs of this interface-driven emission, resulting in an observation that appears to be much more volume-filling.

\subsection{Phase Diagrams}\label{sec:phase}

\begin{figure*} 
    \centering
    \includegraphics[width=1\textwidth]{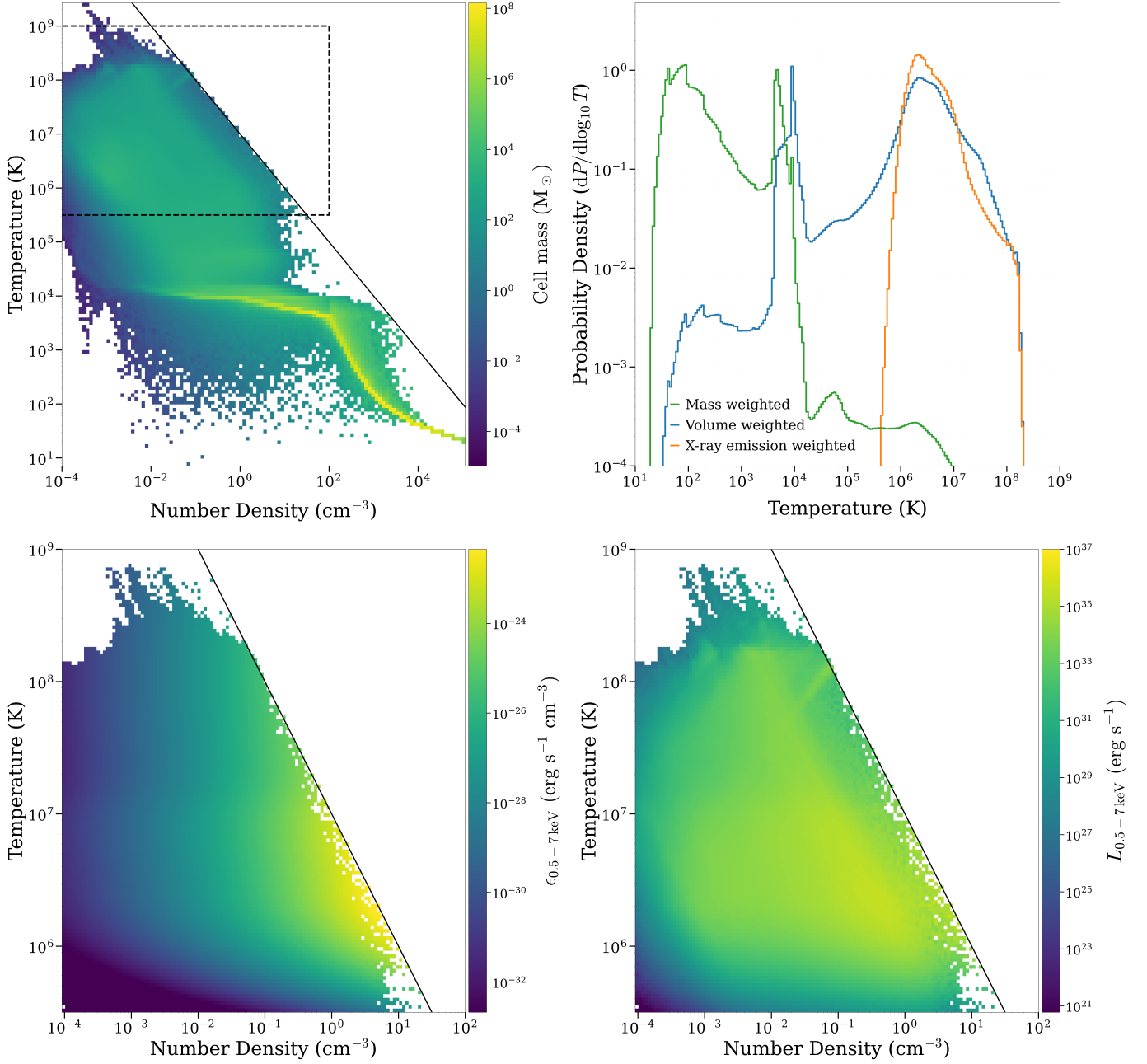}
    \caption{Top left: Density–temperature phase diagram showing total cell mass (M$\odot$) per $(\log{10}\rho,\log_{10}T)$ bin; the dashed box marks the hot-phase region. Top right: Temperature probability distributions weighted by volume (blue), X-ray emission (orange), and mass (green), peaking near $10^{4}$, $2\times10^{6}$, and $10^{2}$ K, tracing the warm, hot, and cold phases. Bottom left: Mass-weighted X-ray emissivity $\epsilon_{0.5-7,\mathrm{keV}}$  in the hot phase, including the outlier bin. Bottom right: X-ray luminosity $L_{0.5-7,\mathrm{keV}}$ summed per bin over the same region. The black solid curve in each phase panel marks the constant-pressure contour $P/k_{\mathrm{B}} = 10^{7}\,\mathrm{K\,cm^{-3}}$, and all cells above this line have been excluded from the analysis as described in the text. Most of the gas mass sits in cold, dense material, yet the X-ray emission comes almost entirely from hot gas at $T \sim (1$--$3) \times 10^6$ K, with the bulk of the luminosity produced at a pressure about an order of magnitude below the peak wind pressure.}

    \label{fig:phase}
\end{figure*}

In Figure \ref{fig:phase} we investigate the phase distribution of gas in our simulation in the temperature-density plane and the contribution of X-ray emission as a function of phase. For the same simulation snapshot as is shown in Figure \ref{fig:slices}, we show the phase diagram weighted by mass (top left), showing the variation of X-ray emissivity across the phase-diagram (i.e. the value of Equation \ref{eq:xray_emissivity} at a given point in the space, bottom left), X-ray emission-weighted phase diagram (essentially the sum Equation \ref{eq:xray_emissivity} for all cells at a certain position in phase space, bottom right) and the temperature distribution of the gas weighted by volume, mass, and X-ray emission.

The temperature distributions clearly distinguish the volume-filling warm and hot phases, the X-ray–bright hot phase, and the mass-dominant, cold component. The X-ray emissivity-weighted distribution peaks at \(T \sim 2 \times 10^6\) K, which is also roughly where the volume-weighted distribution peaks. This indicates that, while X-ray emission is dominated by interfaces between the hottest gas and cooler gas,  these interfaces can be roughly volume-filling. In contrast, the mass-weighted temperature peaks at around \(10^2\) K and shows a secondary maximum near \(10^4\) K, indicating that most of the mass resides in cold, dense gas with substantial contributions from the warm ionized phase; although these components contribute little to the X-ray emission and occupy little volume, they dominate the outflow's mass budget. 

It is more informative to look at the joint 2D $\rho$--$T$ phase diagram.
Due to rare numerical errors in the hydrodynamic solver, occasional dense cells near the injection regions were assigned high temperatures, resulting in single cells with very high pressures and X-ray luminosities. In order to prevent these rare errors from contributing spuriously to the predicted X-ray luminosity, we excised all cells with thermal pressures larger than $P/k_{\mathrm{B}} = 10^{7}\,\mathrm{K\,cm^{-3}}$ which is indicated by the solid lines in each two-dimensional phase diagram.

The X-ray \textit{emissivity} panel (lower-left), showing the mass-weighted mean \(\epsilon_{0.5\text{--}7\,\mathrm{keV}}\), peaks at \(T\approx(1\text{--}3)\times10^{6}\,\mathrm{K}\) and intermediate densities \( n \sim10^{-2}\text{--}10^{2}\,\mathrm{cm^{-3}}\). Unsurprisingly, at fixed temperature, the emissivity increases with increasing density; however, the volume (number of cells) at the highest densities (approaching the constant pressure line shown) is relatively small. This is shown clearly in the \textit{luminosity} panel (lower-right) which shows the contribution from all gas at the given density and temperature. This panel highlights the gas that contributes the bulk of the luminosity, showing clearly that it peaks at a pressure an order of magnitude lower than the maximum wind pressure.

\subsection{Total X-ray luminosity}
\label{sec:timeseries}

\begin{figure}
    \centering
    \includegraphics[width=1\linewidth]{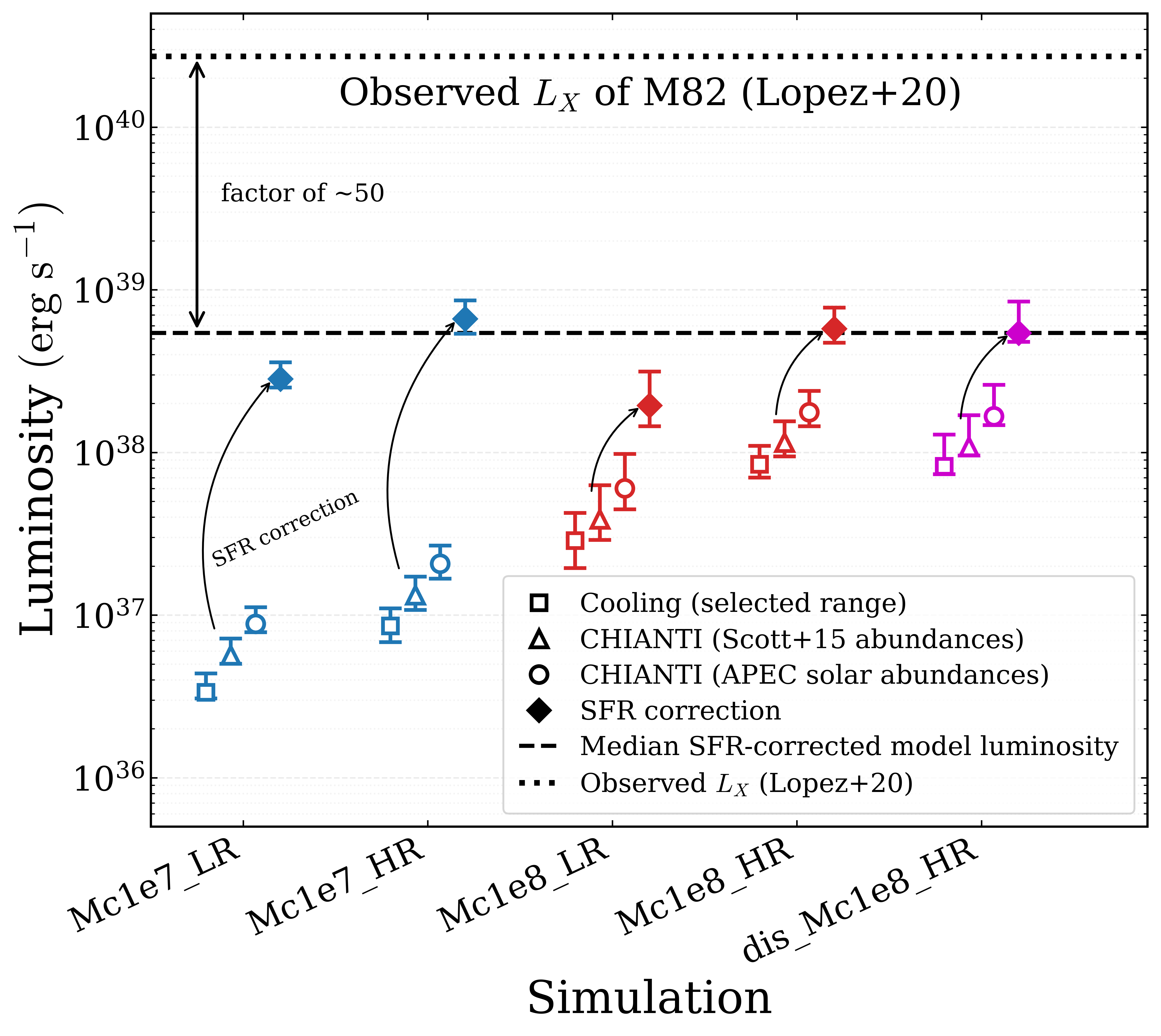}
    \caption{Total X-ray luminosity ($L_X$; 0.5-7 keV) for M82-like outflow simulations, obtained by summing emission over all regions and taking the median across snapshots; error bars show the 16th-84th percentile range. Squares denote cooling in the selected temperature range, triangles show CHIANTI luminosities using \citep{Scott2015} abundances, and circles show CHIANTI luminosities using APEC solar abundances. Filled diamonds indicate CHIANTI luminosities after applying a linear star formation rate (SFR) correction ($\sim50\times$ for $10^7\,M_\odot$ and $\sim5\times$ for $10^8\,M_\odot$ runs), motivated by the simulations linear correlation between input SN energy and $L_X$, which is also seen in observations \citep{Mineo2012}. The dashed line shows the median SFR-corrected luminosity of the simulation \texttt{dis\_Mc1e8\_HR}, while the dotted line marks the observed thermal X-ray luminosity of M82 from \citet{Lopez2020}, after accounting for non-thermal contributions. Arrows indicate the shift from CHIANTI to SFR-corrected luminosities. X-ray luminosity increases with both resolution and cluster mass, yet even the brightest, SFR-corrected simulation falls a factor of $\sim50$ short of the luminosity observed in M82. 
    }
    
    \label{fig:LXstats}
\end{figure}

We next explore how the total X-ray luminosity varies across our simulation suite. Figure \ref{fig:LXstats} shows $L_X$ (0.5$-$7 keV) computed by summing emission over all spatial regions and taking the median across simulation snapshots, with error bars indicating the 16th-84th percentile ranges. We show four different estimates for the total X-ray luminosity. Ordered according to the magnitude of the luminosity that each method produces (as they are ordered from left to right in Figure \ref{fig:LXstats}), these are: (1) the total cooling emission in the simulation from gas with temperatures $k_B T$ in the same energy range as the X-ray band (0.5$-$7 keV)\footnote{We include this for comparison with the results of \citet{Schneider2018,Schneider2020}}, (2) using the \texttt{CHIANTI} code (as described in Section \ref{sec:methods}) with \citet{AndersGrevesse1989} solar abundances, (3) \texttt{CHIANTI} with the \citet{Scott2015} solar abundances, and (4) method (2) with an enhancement ($\sim5\times$ for the $10^8~M_\odot$ runs and $\sim50\times$ for the $10^7~M_\odot$ runs) to account for differences in our effective star formation rate in comparison to the observations (see Section \ref{sec:simulations}).
    
Across all simulations, the CHIANTI-based luminosities computed with different abundance tables show similar trends, with modest normalization differences arising from the adopted abundances. As expected, the three simulations with the largest SN cluster masses ($10^8M_{\odot}$) produce the highest luminosities, reaching median values of roughly $L_X\sim10^{38}-10^{39}$ erg s$^{-1}$ after the SFR-correction. Across the suite, the higher-resolution runs yield systematically brighter emission, suggesting that resolution plays an important role in producing and retaining hot gas. Despite these trends, none of our simulations approach the observed $L_X\approx3\times10^{40}$ erg s$^{-1}$ measured for M82 by \citep{Lopez2020}. Our most optimistic comparison, with the SFR-correction, is still a factor of $\sim \,50$ below the observations, which are indicated with a dotted line in Figure \ref{fig:LXstats}.

\subsection{Mock Chandra X-ray images and spectra}
\label{sec:mocks}

\begin{figure*} 
    \centering
    \includegraphics[width=1\textwidth]{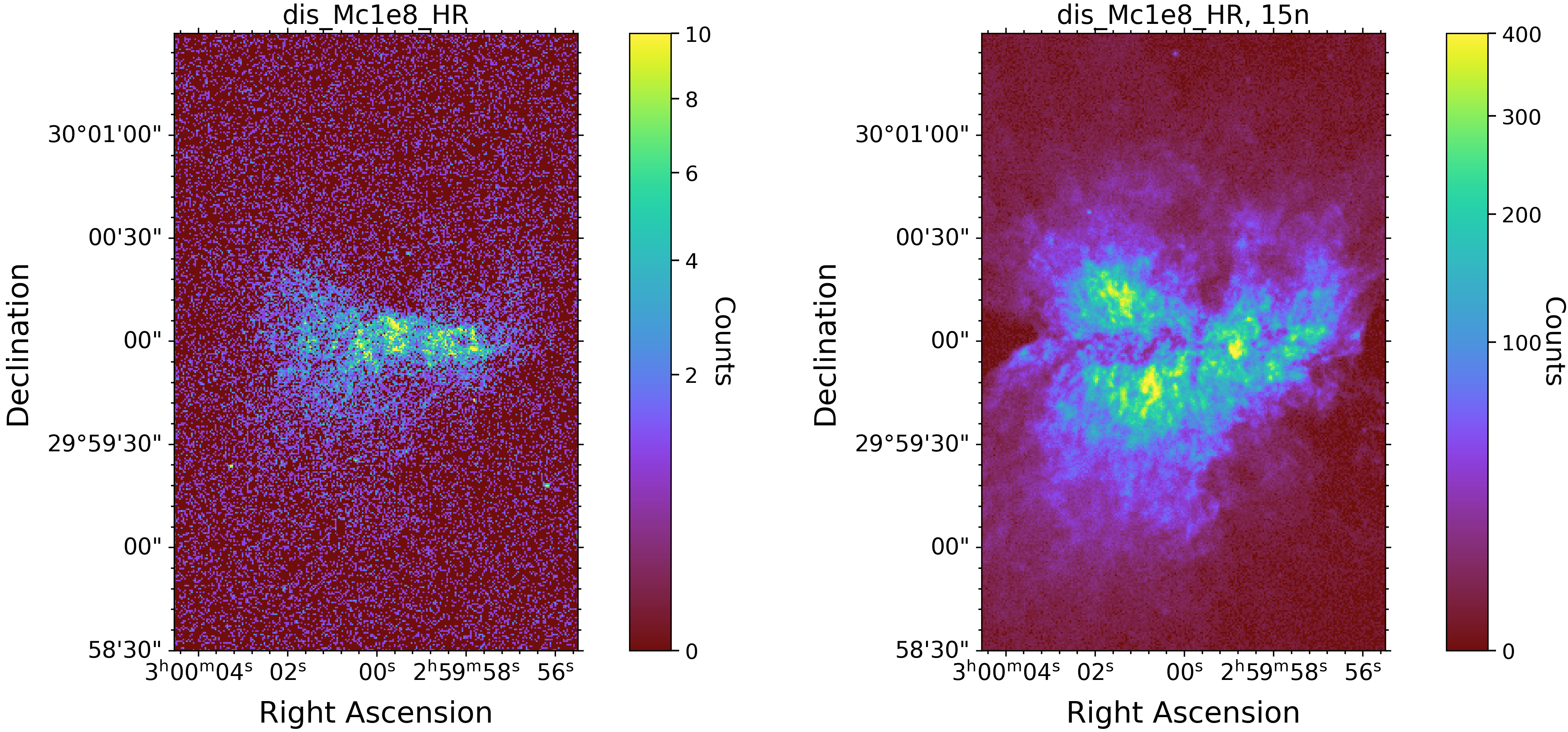}
    \caption{Synthetic X-ray count images generated from \texttt{dis\_Mc1e8\_HR} simulation at t = 90.85 Myr. The left panel shows the fiducial mock Chandra observation, while the right panel shows a rescaled image in which the gas density was artificially increased by a factor $\sim 15$ to boost the X-ray emission. Both images were generated using \texttt{SOXS} with the Chandra ACIS-S, assuming an exposure time of 550~ks and a field of view of $45'' \times 30''$.
    }

    \label{fig:mockimg}
\end{figure*}

Figure \ref{fig:mockimg} presents mock Chandra X-ray images generated with \texttt{SOXS} for the \texttt{dis\_Mc1e8\_HR} simulation at $t=90.85~\mathrm{Myr}$, assuming a $550~\mathrm{ks}$ exposure matched to the M82 observations. The left panel shows the fiducial mock observation, while the right panel shows a visualization test in which the gas density is artificially increased by a factor of $\sim 15$ to boost the X-ray signal to a level comparable to the observed X-ray Luminosity. The fiducial image appears less vertically extended than observed partly because the simulated emission is intrinsically fainter at the matched exposure time. The rescaled image is to illustrate the morphology of the X-ray emission if its luminosity were comparable to the observations.

\begin{figure}
    \includegraphics[width=0.48\textwidth]{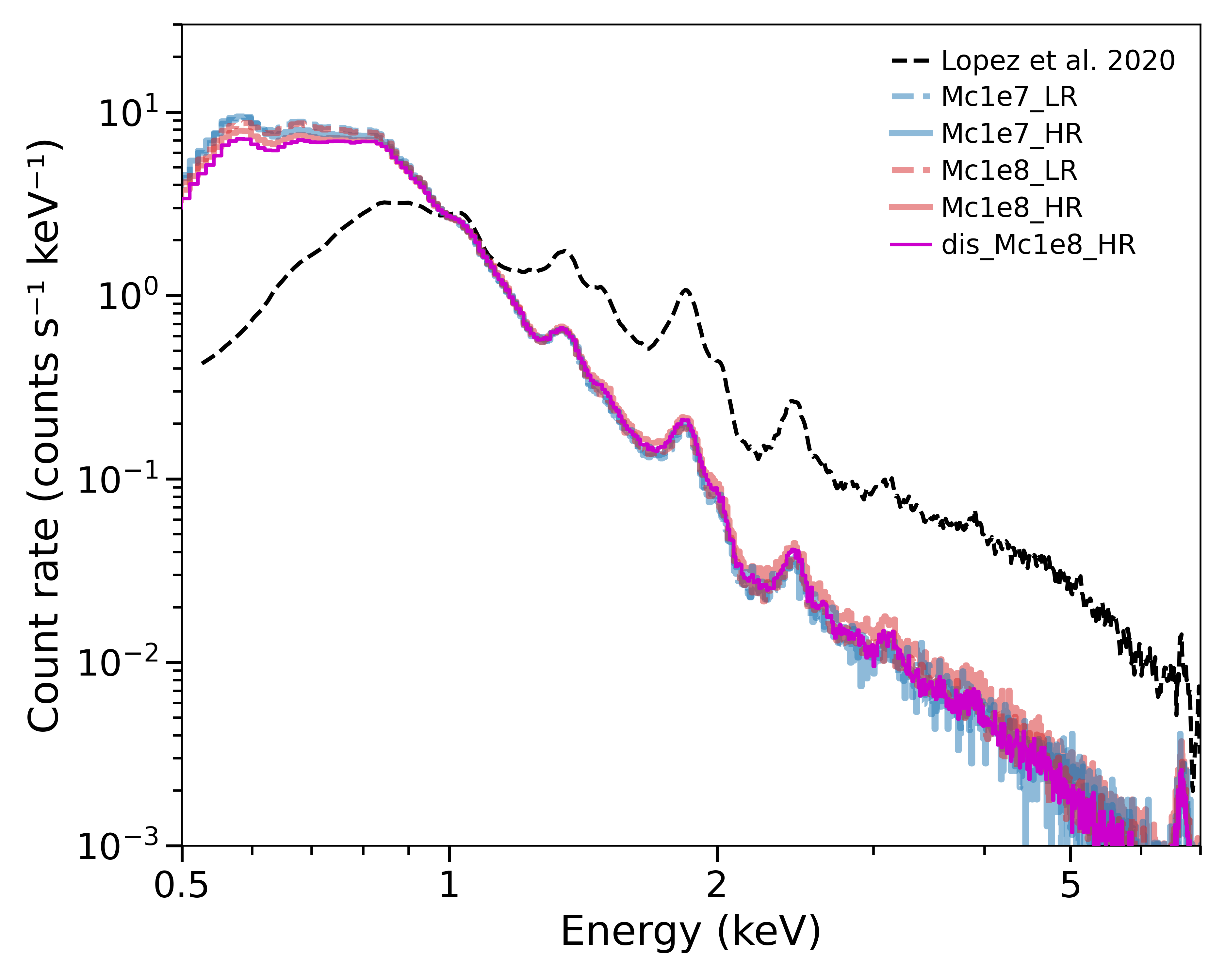}
    \caption{Comparison of simulated X-ray spectra with observations in 0.5–7 keV energy band. The dashed black curve shows the summed observed spectrum from 11 regions (D1, N1-N5, S1-S5; \cite{Lopez2020}). Colored curves show synthetic Chandra spectra from simulations in Table \ref{tab:sim_part1}. All spectra are normalized to match the observed spectrum at 1 keV to emphasize differences in spectral shape. The simulated spectra are systemically too soft, peaking near $\sim 0.55$~keV rather than the observed $\sim 1$~keV, revealing a robust deficit of hard $\gtrsim (1$~keV) photons across all variants.}
    \label{fig:mockspec}
\end{figure}

Figure \ref{fig:mockspec} shows X-ray spectra for each of our simulations compared to the observed spectrum of M82 from \citet{Lopez2020}, with all observational regions combined. Based on Figures \ref{fig:LXstats} and \ref{fig:SB_profs}, we already know that our simulated spectra are much fainter than observed so here we compare the spectral shapes by normalizing each spectrum at 1 keV. The simulated spectra are robustly softer than observed. In order to fit the observed spectrum, \citet{Lopez2020} include a non-thermal component that accounts for $\sim10\%$ of the total $L_X$. Our simulations and mock observations would not account for such emission as we only model thermal emission. We do not expect that adding a power-law contribution will address the discrepancy in spectral shape since this only contributed significantly to emission above 2 keV in \citet{Lopez2020} whereas the discrepancy between our simulated spectrum and the observations is most significant at $0.5-1\,{\rm keV}$. We also tested whether intrinsic absorption within the simulated galaxy could alter the spectral shape by applying the internal absorption functionality in \texttt{pyXSIM} using an approximate neutral hydrogen column density. While this absorption preferentially suppresses emission along high-column sightlines near the disk, it does not substantially change the normalized spectrum. Thus, intrinsic absorption along is unlikely to resolve the spectral hardness discrepancy shown in Figure~\ref{fig:mockspec}.

\subsection{Surface Brightness Profiles}
\label{sec:profiles}

\begin{figure} 
    \centering
    \includegraphics[width=0.49\textwidth]{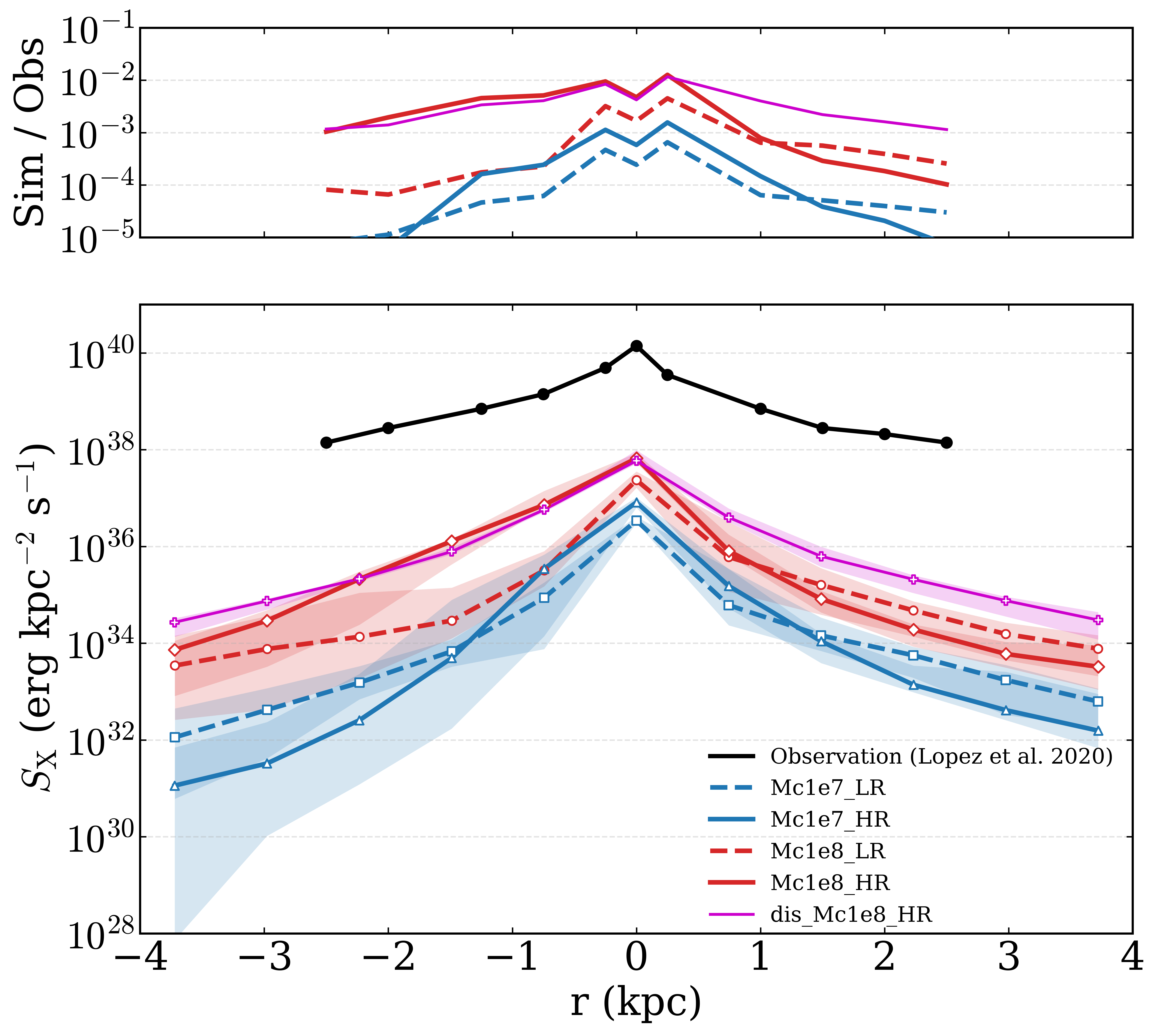}
    \caption{0.5–7.0 keV X-ray surface brightness profiles along the $x$–axis for five simulated M82-like  outflow setups. The profiles are computed using CHIANTI emissivities (CIE, solar metallicity) by integrating along the line of sight in 11 rectangular regions spanning the minor axis, following \citet{Lopez2020}. Style key: $M_{\rm cl}=10^{7},M_\odot$ → blue; $M_{\rm cl}=10^{8},M_\odot$ → red; $\mathrm{dx}=4$ pc → solid; $\mathrm{dx}=8$ pc → dashed; concentrated feedback → thicker line; distributed feedback → thinner line. The shaded regions indicate the 16th–84th percentile range around the median surface brightness for each simulation snapshot. The top panel shows the ratio plot of simulations divided by the observation black line. The observational profile (black line) is transcribed from \citet{Lopez2020}, with the Power-Law and Charge-Exchange (CE) components removed by scaling the observed profile to 70$\%$ of its total value, assuming these components contribute approximately 30$\%$ of the total surface brightness. Our simulated profiles fall off more steeply with height than the observations, with the discrepancy worsening far from the disk.}
 
    \label{fig:SB_profs}
\end{figure}

In Figure \ref{fig:SB_profs}, we compare the median X-ray surface brightness profiles as a function of height above the disk for each simulation against the observed M82 profile from \citet{Lopez2020}. Consistent with the $L_X$ discrepancy discussed above, our brightest simulation is about an order of magnitude fainter than the observational measurement (no correction for SFR has been made here). Even applying the SFR correction discussed in Figure \ref{fig:LXstats} would not bring the simulated profiles into agreement with the observed profile. The discrepancy becomes worse in the outskirts of the profile as our X-ray surface-brightness profiles fall off more quickly than the observations. This is demonstrated by the ratio of profiles given in the top panel of Figure \ref{fig:SB_profs}.

\section{Discussion}
\label{discussion}

In this section, we summarize the main results given in the last section before comparing to previous work and exploring possible reasons for the observed discrepancies between the simulated and observed X-ray properties of M82. In particular, we investigate how the assumption and/or exclusion of different physical processes may contribute to our under-production of X-rays. 

\subsection{Summary of Comparison to Observations}
\label{subsec:obs_comp_sum}

Our simulations reveal a number of discrepancies with the observed X-ray properties of M82, although the robustness of each component varies. We review each in turn below.

\paragraph{Total X-ray Luminosity} The most uncertain comparison is total X-ray luminosity: our simulations fall roughly two orders of magnitude below the observed $L_X \approx 3 \times 10^{40}$ erg s$^{-1}$, but this gap is sensitive to both numerical resolution and the assumed star formation rate. Higher resolution runs produce systematically more emission, potentially because they better resolve the mixing layers which dominate the X-ray emission (see Figure \ref{fig:phase}). Our implied SFR of $\sim 2.5\,M_\odot$ yr$^{-1}$ is a factor of $\sim 5$ below observational estimates for M82. Increasing the SFR by raising the cluster mass injects more energy and mass into the wind, raising the amount of hot gas available to emit in X-rays. Extrapolating to a more realistic SFR---justified by the roughly linear scaling of $L_X$ with injected SN energy both across our simulation suite and in observed galaxies \citep{Mineo2012}---would reduce but not close this deficit. Making the feedback more ``distributed'' also contributes to a higher luminosity, although again, the variation explored here is not enough to resolve the difference.

\paragraph{Surface Brightness Profiles} The second area of disagreement is the surface brightness profile: our simulations broadly reproduce the exponential decline along the minor axis, but, even if we ignore the one to two orders of magnitude difference in normalization, systematically have a steeper drop-off than seen in the data. This shape discrepancy is less sensitive to SFR normalization or resolution and therefore points more directly to missing physics.

\paragraph{Spectral Shape} Our most robust result is the spectral comparison. Normalizing all spectra at 1~keV to isolate variations in the shape of the spectrum (rather than amplitude, which is already probed by our $L_X$ comparison), our simulations consistently produce emission that is too soft, peaking near $\sim 0.55$~keV rather than the observed $\sim 1$~keV. This deficit of hard X-ray photons at $\gtrsim 1$~keV is present across all simulation variants regardless of resolution, feedback strength, or feedback geometry. We also tested that the inclusion of internal absorption produces the same qualitative spectral mismatch, indicating that the hardness discrepancy is not primarily driven by missing intrinsic absorption. Together, these comparisons suggest that while adjustments to the SFR and resolution may partially address the luminosity deficit, the spectral hardness and extended surface brightness profile likely require additional physical ingredients not yet captured in our models.

\subsection{Comparison to previous work}
\label{subsec:prev_work}

We are not the first to simulate the M82 wind. We first note that variants of these same simulations were presented in \citet{TanBretFielding2024} and shown there to agree well with observations of other outflow components such as cold clouds, demonstrating that clustered supernova-driven winds produce long-lived cold gas structures whose sizes and mass evolution are consistent with observational constraints. The agreements with observations in those phases, contrasted with the discrepancies in the X-ray properties here, emphasizes the need for multiwavelength constraints when investigating the physics of galactic winds.

Several previous studies have modeled X-ray emission in star-forming galaxies and galactic winds in a wide range of contexts. \citet{Strickland2000} carried out an extensive hydrodynamical parameter study of starburst-driven winds motivated by M82. Consistent with the simulations presented here, \citet{Strickland2000} showed that the soft X-ray emission arises from low-filling factor gas that contains only a small fraction of the wind's mass and energy, with the emission originating primarily from mixing layers between the hot wind and cooler ambient gas, while most of the energy resides in the the low-density, hot, volume-filling phase that is difficult to observe directly. \citet{Cooper2008} used three-dimensional simulations with an inhomogeneous ISM and found that soft X-ray emission is closely associated with shocked gas and interactions between the hot wind and entrained cool material.

\citet{Sarkar2016} modeled diffuse X-ray emission using two-dimensional hydrodynamical simulations of supernova-driven outflows and \citet{Chevalier&Clegg1985}-inspired models with an SFR-dependent mass-loading parameter, $\beta$. They show that soft X-ray emission is dominated by the free-wind phase in their simulations, justifying a \citet{Chevalier&Clegg1985}-like model for X-ray emission, but standing in contrast to the present work and other three-dimensional simulations which demonstrate mixing layers as the dominant contribution to X-ray emission \citep{Strickland2000,Cooper2008,Vijayan2018,Schneider2024,Huong2025}. In the context of their models they show that the slightly super-linear relationship between SFR and $L_X$ \citep{Wang16} can be re-produced by their SFR-dependent mass-loading, which they justify on physical grounds. In follow-up work, \citet{Vijayan2018} used three-dimensional simulations of disc-wide supernova injection to show that extra-planar X-ray emission likely arises from bow-shock and mixing regions surrounding warm clumps lifted from the disc. This is consistent with the simulations shown here as well as the clear extra-planar emission seen in the observations \citep{Lopez2020}, which would not be explained by a \citet{Chevalier&Clegg1985}-like model with centrally concentrated X-ray emission. This emphasizes the need for multiple points of comparison with observations in order to distinguish between different physical scenarios. A study by \citet{Jana2024} showed how strong galactic outflows interact with extra-planar and circumgalactic gas using controlled hydrodynamical simulations, showing that the soft X-ray response depends on the energy injection rate and on the initial thermodynamic structure of the CGM. 

\citet{Huong2025} use the QED simulations \citep{QEDI,QEDII}, based on the GPU-accelerated \texttt{QUOKKA} code \citep{QuokkaI}, to demonstrate that the evolution of metallicity with galactic height as observed in observations of \citet{Lopez2020} can be explained through mixing of SNe material with swept-up ISM material. This results in lower-metallicity emission further from the disc, as is observed. This result again emphasizes the importance of mixing layers to explain observed X-ray emission. \citet{Huong2025} is the only other work to produce mock X-ray spectra from their simulations (they also use \texttt{pyXSIM} and \texttt{SOXS}), but they do not provide quantitative comparison of this spectrum to the observations as the underlying simulations were not made to be representative of M82.

\citet{Schneider2018}, \citet{Schneider2020}, and \citet{Schneider2024} carried out high-resolution ($\lesssim$5 pc) global simulations of an isolated, M82-like galaxy using the GPU-accelerated code \texttt{Cholla} as part of the CGOLS project. \citet{Schneider2018} argue that while their simulations are converged in soft X-ray emission, even slightly lower resolution simulations ($\Delta x\sim 10\,{\rm pc}$) are not converged, consistent with our two values of resolution shown here ($4$ \& $8\,{\rm pc}$) which vary in X-ray emission. This emphasizes that extreme resolution is required to fully capture the multiphase structure of galactic winds relevant for X-ray emission. They find that their simulation was roughly able to reproduce the X-ray surface brightness profile of M82 from \citet{Strickland2004}, though it does fall off moderately more quickly \citep[see][Figure 10]{Schneider2018}. \citet{Schneider2024} extend this work to include a suite of simulations that also model spatially extended star formation feedback, which they demonstrate tends to increase the dominance of the cooler phase leading to more radiative winds and slightly less extended X-ray emission. There are three key differences between our simulations and the CGOLS suite that could lead to discrepancies in estimated X-ray emission. First, in the calculation of $L_X$ used in \citet{Schneider2018} and \citet{Schneider2024}, the authors assume X-ray emissivity to be given by the total cooling rate of all gas cells with temperatures within the corresponding X-ray band (e.g. $kT$ between 0.3 and 2.0 keV). We perform a similar calculation in Figure \ref{fig:LXstats} and Appendix \ref{sec:appendix} and show that, at least for our adopted cooling function, this does not result in significant differences in $L_X$ estimates. Second, both CGOLS papers assume a significant period of the simulation with SFR of 20 $M_{\odot}$/yr which is a factor of ten higher than ours and a factor of $\sim 2$ greater than estimates for M82 \citep[$\sim13$ $M_{\odot}$/yr;][]{forsterschreiber03}. This higher adopted SFR could (partially) help to explain the fact that the CGOLS simulations give $L_X$ consistent with the observations while ours are too low. Accounting for this SFR in our estimates, by, for example, multiplying our estimated $L_X$ by a factor of $\sim 10$ for the \texttt{Mc1e8} simulations, would still result in a factor of $\sim 25$ discrepancy with the observations. Third, the CGOLS simulations have a minimum gas temperature of $10^4\,{\rm K}$, resulting in lower density ISM clouds which are potentially more easily shredded and mass-loaded into the galactic wind. This could result in higher X-ray emission in the mass-loaded warm-hot phase $\sim 10^6 \, {\rm K}$). Our simulations are of comparable spatial resolution and capture the vast majority of the X-ray emitting region contained in the CGOLS works. We therefore posit that the discrepancy in estimated X-ray emission originates either from the differences in ISM phase treatment (minimum temperature), as outlined above, differences in the adopted cooling function, which would affect their X-ray emission estimate, or differences in how the supernovae feedback is injected (either in terms of the thermodynamic properties of the gas or its spatial/temporal injection).

In more recent work, \citet{Li2025} performed three-dimensional hydrodynamical simulations to study the multiphase galactic winds in M82 using the \texttt{Athena++} code at a resolution of $\sim$10 pc. Unlike our tall-box setup, their simulations span 8 kpc in all three dimensions. A key ingredient in their work is the implementation of ``gas return'' from sink particles to account for the fact that only a small fraction of gas within giant molecular clouds (which the sink particles approximately represent) is converted into stars. The gas return prescription of \citet{Li2025} assumes that some of the gas in sink particle remains as gas and is slowly returned to the surroundings (the grid) with a given efficiency and timescale, which they vary. This prescription results in effectively higher mass-loading of the SN-driven wind, as more mass is injected per unit SN. In their models without gas return, the ratio of simulated to observed $L_X$ is about 14\%-26\%, which is closer to, but slightly larger than, our work. In contrast, when including gas return, their $L_X$ rises to ~60\%-150\% of the value measured by \citet{Lopez2020}, while increasing the initial gas mass alone moderately enhances X-ray luminosity but is insufficient to match observations. This suggests that mass loading of the recycled gas is important for increasing X-ray emission, consistent with \citet{Sarkar2016}.

\subsection{Relevant missing physical processes}
\label{relevant missing physical processes}

Our simulations fall short of the observed X-ray luminosity and produce spectra that are too soft across all simulation variants. We explore here how physical processes missing from our simulation could help explain these discrepancies. 

\subsubsection{Thermal conduction}
\label{subsubsec:conduction}

Thermal conduction at the interfaces between hot and cold phases can drive mass loading of the hot phase through evaporation of cold clouds \citep{Cowie1977}, which in turn enhances radiative emission from the hot gas \citep{Weaver1977, EL-Badry2019}. This process could increase the density of the hot, X-ray emitting plasma and boost emission at higher temperatures, potentially explaining the deficit of hard X-ray photons between our simulations and the observations, while simultaneously increasing the X-ray luminosity.

This interpretation is consistent with a recent work on wind-blown bubbles in the Tarantula Nebula in the LMC \citep{Rodriguez2026}, which found that simulations similarly under-produce hard X-rays and under-predict emission from bubble interiors relative to the observations. Taken together, these results suggest that thermal conduction may be a generic missing ingredient in multi-phase wind simulations. Future work incorporating this process could help to explain the X-ray properties of M82. 

\subsubsection{Cosmic rays}
\label{subsubsec:crs}

Cosmic rays (CRs) are relativistic charged particles -- produced primarily by supernova shocks -- that contribute an additional pressure component on top of the thermal pressure of the gas. They are expected to have a profound effect on the thermal structure of galactic winds \citep{Hopkins2020}, in particular making the CGM cooler and denser \citep{Ji2020}, though these results depend sensitively on assumptions around CR transport \citep{SalemBryan2014, Chan2019}. In particular, CRs could help produce more of the warm-hot ionized medium ($T \sim 10^{5.5}$--$10^7$ K) that efficiently emits X-rays through heating via the dissipation of Alfv\'{e}n waves \citep{Armillotta24,Kim2026}.

Additionally, inverse Compton scattering -- in which low energy CMB photons are up-scattered to X-ray energies by relativistic CR electrons -- has recently been used to explain X-ray emission in cool-core clusters and the CGM \citep{Hopkinsa2025,Hopkinsb2025}. While this process could help explain the power-law like tails at high energies in the observed spectrum of M82 (see Figure \ref{fig:mockspec}), the presence of clear line emission -- for example SXV emission above the disk \citep[Figure 4]{Lopez2020} -- requires significant emission from gas at $T \sim 10^7$ K that is missing in our simulation. Inverse Compton scattering alone therefore seems unlikely to account for the full discrepancy in X-ray emission discussed in this work. 

\section{Conclusion}
\label{conclusion}

We have carried out high-resolution ($\sim4-8$ pc) ``tall box'' ($\sim2\times2\times8$ kpc$^3$) MHD simulations of an M82-like starburst galaxy to investigate the structure and X-ray emission of its galactic wind. Using synthetic observations generated with \texttt{pyXSIM} \citep{ZuHone2016} and \texttt{SOXS} \citep{ZuHone2023}, we compared the simulated X-ray surface brightness profiles, spectra, and total luminosities to deep Chandra data from \citet{Lopez2020}. We summarize key findings as follows:

\begin{itemize}
\item Temperature--density phase diagrams confirm that while most of the gas mass resides in cold, dense gas, the X-ray emission originates from a small fraction of hot plasma at $T \sim (1$--$3) \times 10^6$ K representing a relatively small volume near the interfaces of hot gas with cooler clumps entrained in the wind. The bulk of the X-ray luminosity is produced at pressures an order of magnitude below the peak wind pressure (Figure~\ref{fig:phase}).

\item The total predicted X-ray luminosity increases with both resolution and starburst cluster mass, and is higher when SNe are distributed throughout the disk rather than concentrated in the nucleus. However, even our highest-luminosity simulation falls nearly two orders of magnitude below the observed $L_X \approx 3 \times 10^{40}$ erg s$^{-1}$, a gap that is only partially explained by our relatively low assumed SFR of $\sim 2.5\,M_\odot$ yr$^{-1}$ (Figure~\ref{fig:LXstats}).

\item The simulated X-ray spectrum is consistently too soft across all simulation variants, regardless of resolution or feedback geometry. Normalizing at 1~keV to isolate spectral shape, our emission peaks near $\sim 0.55$~keV compared to $\sim 1$~keV in the data, indicating a robust deficit of hard X-ray photons at $\gtrsim 1$~keV that cannot be attributed to the missing non-thermal power-law component (Figure~\ref{fig:mockspec}).

\item Our simulations broadly reproduce the exponential decline of X-ray surface brightness along the minor axis, but generally decline more steeply with distance from the disk. The distributed feedback model performs best, though still falls well short of the data (Figure~\ref{fig:SB_profs}).

\end{itemize}

Taken together, these results highlight both the promise and the limitations of current high-resolution idealized simulations in capturing the X-ray properties of starburst-driven galactic winds. The luminosity deficit is the most uncertain of these discrepancies, as it is sensitive to the assumed SFR and numerical resolution, while the spectral shape mismatch is the most robust, persisting across all simulation variants. The surface brightness profile discrepancy lies in between, reflecting both the normalization issue but also a deficit in spatially extended hot gas. The discrepancies likely reflect the absence of key physical processes including thermal conduction, cosmic ray feedback, and approximations in our treatment of supernovae feedback and the entrainment of gas near star forming regions. Inclusion of these physical processes in future work, along with comparison to multi-wavelength tracers \citep[e.g.][]{Lopez2025} is essential for enabling a full accounting of the physics at play in the driving of galactic winds.

\section*{Acknowledgements}

AL and VP were supported by NSF Astronomy \& Astrophysics grant No.2307419. VP was also supported by the NASA Hubble Fellowship grant HST-HF2-51489 awarded by the Space Telescope Science Institute, which is operated by the Association of Universities for Research in Astronomy, Inc., for NASA, under contract NAS5-26555. D.B.F. gratefully acknowledges support from NSF through grants AST-2407387 and from NASA through grants HST-AR-17859.015-A and HST-AR-17559.009-A. GLB acknowledges support from the NSF (AST-2108470, AST-2307419), NASA TCAN award 80NSSC21K1053, and the Simons Foundation through the Learning the Universe Collaboration. SL and LAL were supported by NASA’s Astrophysics Data Analysis Program under grant No. 80NSSC22K0496SL, and LAL also acknowledges support through the Heising-Simons Foundation grant 2022-3533.

\appendix

\section{Variations in Theoretical X-ray Emission Calculations}
\label{sec:appendix}

To predict X-ray emission from our simulations, we employ two independent methods as described in Section \ref{sec:methods}: a direct emissivity lookup using pre-tabulated CHIANTI tables, and a Monte Carlo synthetic observations pipeline using \texttt{pyXSIM} and \texttt{SOXS}. Here, we validate that these two approaches yield consistent emissivities, and we quantify the sensitivity of our results to the choice of collisional ionization equilibrium (CIE) model. 

In Figure \ref{fig:radiative_cooling_xray} we compare variations in X-ray emissivity that can originate from differences in adopted abundance patterns and provide a comparison against the emissivity calculation of \citet{Schneider2018}. In the left panel, we show $\Lambda(T) \equiv \dot{\epsilon}(T)/n_\mathrm{tot}^{2}$ as a function of gas temperature, where $\Lambda(T)$ represents different emissivity functions. The light orange curve shows the total broadband radiative cooling function employed in our simulations. The method of \citet{Schneider2018} calculates X-ray emission using all cells between the two black lines, corresponding to temperatures $kT$ between $0.5-7\,{\rm keV}$. We use this method for the open squares shown in Figure \ref{fig:LXstats}.

The blue curve in Figure \ref{fig:radiative_cooling_xray} shows the $0.5$--$7~\mathrm{keV}$ band emissivity calculated from \texttt{CHIANTI} using its default solar abundances \citep{Scott2015} whereas the red curve shows this same band emissivity, also calculated with \texttt{CHIANTI} but now using the default abundance pattern used in \texttt{pyXSIM} \citep{AndersGrevesse1989}. The two X-ray emissivity curves agree well across the temperature range most relevant to X-ray emission ($T \sim 10^{6}$--$10^{8}$~K), though differ by a factor of $\sim 2$ at temperatures of $\sim 10^7 \, {\rm K}$. The right panel explains why this is by comparing the abundance tables used by each calculation: differences in the adopted abundances, particularly for Carbon, Nitrogen, Oxygen, and Iron ($Z = 6,\,7,\,8,\, \& \, 26$ respectively). We find that the total X-ray emissions are 3.40$\times$10$^{38}$ ergs/s and 1.64$\times$10$^{38}$ ergs/s for the \citet{AndersGrevesse1989} (labeled as \texttt{pyXSIM}) and \citet{Scott2015} (labeled as CHIANTI) abundances, respectively for \texttt{dis\_Mc1e8\_HR}.

We also tested fluctuations in the X-ray emission calculated using different Collisional Ionization Equilibrium (CIE) models available in \texttt{pyXSIM}, shown in Table \ref{table: different models SOXS}. Specifically, the APEC and SPEX models are widely used in X-ray astronomy and include detailed line emission from elements such as C, N, O, Ne, Mg, Si, S, Ar, Ca, Fe, and Ni, with other elements fixed at solar abundances. The Cloudy model, on the other hand, provides a broader treatment of chemical species, including all elements up to Zn, but differs somewhat in its emissivity predictions compared to APEC and SPEX; see \texttt{pyXSIM} documentation for details \citep{ZuHone2016}.

\begin{figure}[H]
    \centering
    \includegraphics[width=1\textwidth]{cooling_comparison.png}
    \caption{Left: Radiative cooling function as a function of gas temperature. The light orange curve shows the total broadband radiative cooling function used in our simulations. The blue and red curves shows the $0.5$--$7~\mathrm{keV}$ band emissivity computed from \texttt{CHIANTI} using the \citet{AndersGrevesse1989} (red) and \citet{Scott2015} (blue) solar abundance patterns. Right: Solar photospheric abundances for \citet{Scott2015} (open blue squares) and \citet{AndersGrevesse1989} (open red triangles), plotted against atomic number $Z$ versus  $12+\log_{10}(n_X/n_H)$.}
    
    \label{fig:radiative_cooling_xray}
\end{figure}

\begin{table}[H]
\centering
\begin{tabular}{lc}
\toprule
\textbf{CIE Models (\texttt{SOXS})} & \textbf{X-ray Emission} \\
\midrule
\texttt{apec}    & 3.39$\times$10$^{38}$ erg s$^{-1}$ \\
\texttt{spex}    & 3.50$\times$10$^{38}$ erg s$^{-1}$ \\
\texttt{cloudy}  & 2.79$\times$10$^{38}$ erg s$^{-1}$ \\
\bottomrule
\end{tabular}
\caption{X-ray emission calculated values using different collisional ionization equilibrium (CIE) models from \texttt{SOXS}. For this comparison, emissivity values were computed over the energy band 0.5--7 keV, assuming a gas metallicity of $Z_\mathrm{gas} = 1Z_\odot$. See \texttt{pyXSIM} documentation \citet{ZuHone2016} for details on each models. }
\label{table: different models SOXS}
\end{table}

\bibliography{sample631}{}
\bibliographystyle{aasjournal}

\end{document}